\documentclass[aps,preprint,unsortedaddress]{revtex4-1}
\usepackage[utf8x]{inputenc}
\usepackage{graphicx}
\usepackage{amsmath, amssymb, setspace}
\usepackage{epstopdf}
\usepackage{color}
\usepackage{bm}% bold math
\usepackage{subfigure}
\usepackage{subfloat}

\newcommand{\Ca}{$^{40}$Ca$^{+}$}

\newcommand{\Ar}{Ar$^{+}$}

\begin{document}

\title{Probing surface electric field noise with a single ion}

\author{N. Daniilidis$^{1,*}$, S. Gerber$^{1,*}$, G. Bolloten$^{1}$, M. Ramm$^{1}$, A. Ransford$^{1}$, E. Ulin-Avila$^{1}$, I. Talukdar$^{1}$, and H. H\"affner$^{1,2}$}

\affiliation{$^1$Department of Physics, University of California, Berkeley, 94720, Berkeley, CA}
\affiliation{$^2$Materials Sciences Division, Lawrence Berkeley National Laboratory, Berkeley, CA $94720$}

\thanks{These authors contributed equally to this work.}

\email{email correspondence to: hhaeffner@berkeley.edu}

\date{\today}

\begin{abstract}

We report room-temperature electric field noise measurements combined with \emph{in-situ} surface characterization and cleaning of a microfabricated ion trap. 
We used a single-ion electric field noise sensor in combination with surface cleaning and analysis tools, to investigate the relationship between electric field 
noise from metal surfaces in vacuum and the composition of the surface. These experiments were performed in a novel setup that integrates ion trapping capabilities 
with surface analysis tools. We find that surface cleaning of an aluminum-copper surface significantly reduces the level of electric field noise, but the surface 
does not need to be atomically clean to show noise levels comparable to those of the best cryogenic traps. The post-cleaning noise levels are low enough to allow 
fault-tolerant trapped-ion quantum information processing on a microfabricated surface trap.

\end{abstract}

\maketitle

Noise and dissipation near surfaces and interfaces present challenges in many fields of science and technology. This includes modern nanoelectronics \cite{Simoen2012,Ishigami2006}, 
superconducting electronics \cite{Pashkin2009}, studies of non-contact friction \cite{Stipe2001}, microtraps for ions \cite{Hite2012} and ultracold atoms \cite{Henkel1999}, 
detection of Casimir forces \cite{Kim2010a}, and tests of general relativity \cite{Pollack2008, Everitt2011}. It is, therefore, imperative to gain a better understanding 
of the sources of such noise, so that appropriate solutions can be adopted. In particular, electric field noise near the electrode surfaces of ion traps has been known 
to cause unexpectedly high heating of the motional modes of a trapped ion \cite{Wineland1998}. Recently, a 100-fold reduction of the noise level was achieved after sputter 
cleaning of a trap surface \cite{Hite2012}. The noise seems to have been caused by carbon contamination of the gold surface, but the actual mechanisms are not yet understood. 
While the noise reduction is a major advance towards scalable trapped-ion based quantum computing, such work also shows that trapped ions can serve as ultra-sensitive detectors 
of certain surface properties.

Detectors operating in the quantum regime can offer unparalleled levels of sensitivity \cite{Clerk2010}. Such systems can be used to measure weak forces \cite{Teufel2009,Maiwald2009a}, 
magnetic fields \cite{Budker2007}, and charges \cite{Devoret2000}. In the case of trapped ions, experimentalists have achieved extremely accurate control in preparing and measuring 
the ion quantum state \cite{Haeffner2008}. Thus, by measuring the effect of electric field noise on the motional quantum state of the ion, one can probe the noise in the frequency 
range from 100~kHz to few MHz with remarkable sensitivity \cite{Wineland1998,Maiwald2009a}. Such studies will ideally be performed in combination with other surface-characterization 
and modification tools to provide complementary information about the surface.

Here we combine, for the first time, a single-ion noise sensor in the same apparatus with surface analysis tools. We use single \Ca trapped ions to measure electric field noise 
arising from the trap electrodes, sputter clean the electrode surfaces using an \Ar~ion beam, and correlate the noise spectra with the surface composition determined using \emph{in-situ} 
Auger spectroscopy. After cleaning, the energy relaxation time for the motional state of a \Ca ion trapped at 1~MHz is $\tau_1\approx$260~ms, long enough to allow fault-tolerant quantum 
information processing \cite{Haeffner2008}. We find that the surface need not be atomically clean to show such low noise levels. Measurements performed with a residual gas analyzer 
suggest that the hydrocarbon molecules on the surfaces of our vacuum system prior to surface cleaning are generally heavier than those present after cleaning.

\begin{figure}
\begin{center}
\includegraphics[width = 0.4\textwidth]{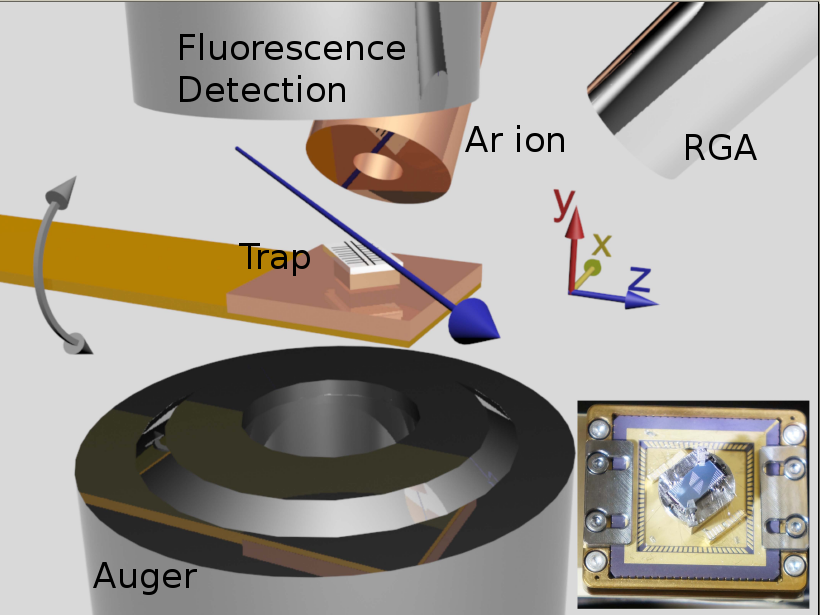}
\caption{\label{setup} The vacuum system integrates a surface trap (center) attached to a filter board (yellow, left) on a 360$^{\rm o}$ rotational holder (rotation about the $z$ axis), 
an Auger spectrometer (bottom), an Ar$^+$ gun (Ar ion, behind $xy$ plane), an observation channel (Fluorescence Detection, above),  and a residual gas analyzer (RGA, upper right). The 
laser direction (blue arrow) is at a 7$^{\rm o}$ angle with the trap axis, defined by the radio-frequency trap electrodes (black lines on the trap). A photograph of the mounted trap 
is shown in the lower right corner (color online).}
\label{fig:sphere}
\end{center}
\end{figure}

We performed the experiments in a multi-function vacuum system consisting of a 12 inch spherical chamber, which allows surface treatment, characterization, and ion trap operation, without 
breaking vacuum. We show a schematic of the experimental setup in Fig.~\ref{fig:sphere}.  A microfabricated ion trap is mounted on a manipulator and can be rotated around the z axis. 
Depending on the trap orientation, the trap chip is facing a retractable viewport with a large NA (approximately 0.3) objective for ion fluorescence collection ($+\mathbf{y}$ direction), 
an \Ar~ion gun (45$^{\rm o}$ with respect to the $\mathbf{y}$ axis), or an  Auger/LEED unit ($-\mathbf{y}$ direction), which allows us to characterize the elemental composition of the 
trap surface. A residual gas analyzer (RGA) is installed in the same chamber, with its axis at 45$^{\rm o}$ with respect to $\mathbf{y}$ on the $yz$ plane and allows us to analyze the 
residual gases in the chamber, as well as the outgassing from adsorbate layers forming on the chamber walls. Passive low-pass filters and shielding help reduce the electronic noise 
reaching the trap electrodes (see App.~\ref{app:A}).

In this work, we used a microfabricated surface electrode trap, trapping ions at a distance of 100~$\mu$m from the nearest electrode. The trap consists of a fused quartz chip on which 
the electrode pattern was etched using a combination of excimer-laser weakening and HF-etching (performed by Translume, Ann Arbor, MI). After the substrate was etched, we cleaned it 
in Piranha solution at 120$^{\rm o}$C, and evaporated a metal film combination of 15~nm Ti, 500~nm Al, 30~nm Cu, 15~nm Ti, 500~nm Al, 30~nm Cu, in an e-beam evaporator with vacuum better 
than $3\times10^{-6}$~mbar, without allowing the aluminum surfaces to oxidize in air before the copper layer evaporations. The copper layers serve the purpose of preventing oxidation of 
the aluminum surfaces upon exposure to air. Subsequently, we mounted the trap on a chip carrier without allowing it to come in contact with any solvents, and baked it in the vacuum system at 160$^{\rm o}$C for three weeks, to achieve 
vacuum of approximately 10$^{-10}$~mbar.

After baking, we used single ions to measure the electric field noise of the trap and  performed surface analysis using the Auger spectrometer. Following this, we 
cleaned the trap surface under an \Ar~beam and re-analyzed the surface after cleaning. We then used a single ion to measure the noise spectrum after a 40 day waiting period, and recorded 
Auger spectra from the surface within a few days from the noise measurement. Finally, we repeated the surface cleaning, noise measurement, and Auger analysis steps one more time, all within 
a few days from each other. We performed both \Ar cleaning runs under the same conditions. The argon pressure was $10^{-4}$~mbar, the ion beam energy was 300~eV, the angle of incidence was 
perpendicular to the trap surface, and the beam flux was $2\times10^{17}\tfrac{1}{{\rm m}^2\,{\rm s}}$. We carried out each cleaning step for a total of 20 minutes, resulting in an estimated 
removal of 10~nm of material from the surface.

\begin{figure}
%\begin{center}
\includegraphics[width = 0.4\textwidth]{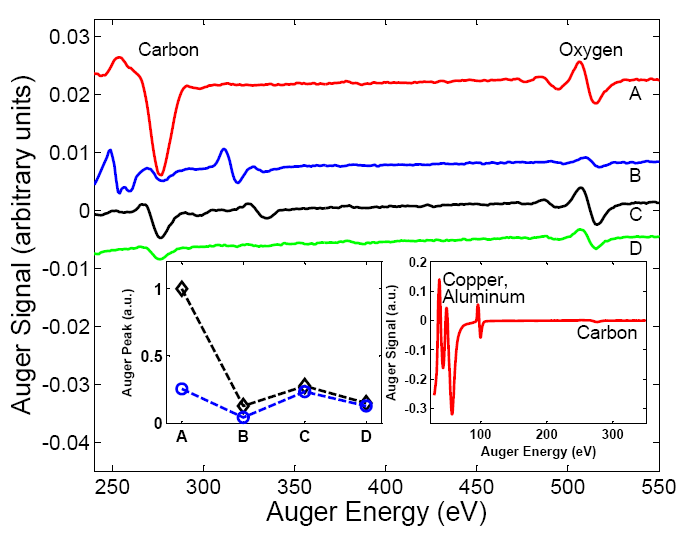}(a)
\noindent
         \begin{tabular}{|c||c|c|c|c||c|c|c|c|}
         \hline
         (b)& A$_{(1)}$ & B$_{(1)}$ & C$_{(1)}$ & D$_{(1)}$ & A$_{(2)}$ & B$_{(2)}$ & C$_{(2)}$ & D$_{(2)}$ \\
         \hline
         O monolayers &0.02&$0.002$&0.02&0.01 &  0.8 & 0.3 &1.1 &  0.7\\
         \hline
         C monolayers &0.08&$0.01$&0.03&0.02 & 0.6&0.1&0.3&0.2\\
         \hline
         \end{tabular}
\includegraphics[width = 0.38\textwidth] {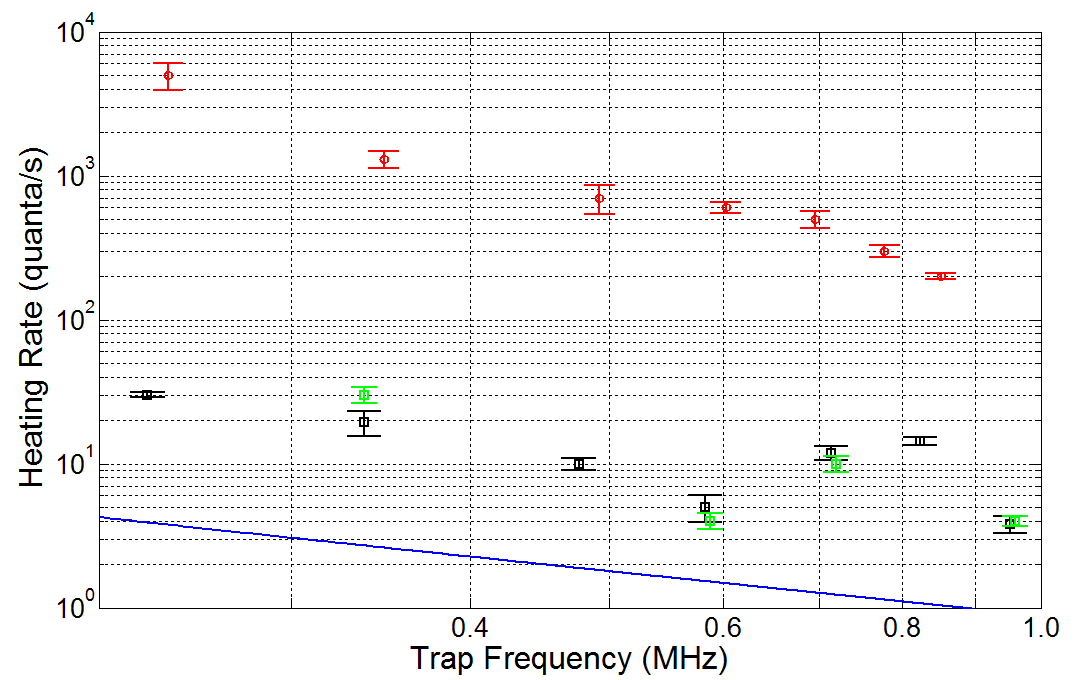}(c)
\caption{(a) Auger spectra at different stages of surface treatments: After the bake (red, A). After the first clean (blue, B). After 40 days in vacuum (black, C). After the second clean 
(green, D). In addition to carbon and oxygen, a palladium peak, coming from the Pd wirebonds used, appears at 330~eV in B and C. The inset on the left plots the evolution of the signal 
strengths of oxygen (blue, circle) and carbon (black, diamond) normalized to the pre-clean carbon value.  The inset on the right shows a low energy Auger spectrum after the vacuum bake, 
with peaks corresponding to oxidized Al(LMM), Cu(MVV) and C(KLL). (b) Oxygen and carbon coverage of the trap surface, at different stages. Left and right sets correspond to different 
sensitivity calibration scales of the Auger electron analyzer (see text). (c) Frequency scaling of motional heating in our trap. Color coding for measured data in (a) and (c) is the same 
(color online). Solid line shows the expected Johnson noise heating for our setup. The trapping height is 100~$\mu$m. } \label{data}
%\end{center}
\end{figure}

In Fig.~\ref{data}a, we show the Auger spectra of the trap surface at various stages, taken with incident electron energy of 2~keV. 
After baking, both carbon and oxygen were present on the trap surface. After cleaning, the carbon and oxygen KLL peaks were reduced by approximately 90\%, but their ratio remained unchanged 
(see \cite{Davis1978} for peak nomenclature and identification). In addition, the chemical shift of the Al(LMM) peak indicates that the aluminum oxide was removed. After the wait period 
of 40 days, the carbon peak increased roughly by a factor 2, whereas the oxygen peak increased by a factor of 5, and the aluminum partially oxidized. Thus, in this period carbon compounds 
were re-adsorbed or diffused to the surface, and some oxidation of the trap surface took place. After the second cleaning step, both carbon and oxygen peaks were reduced by a factor of 2, 
but some oxide remained.

We use the strengths of the copper MVV peak, aluminum and aluminum oxide LMM peak, carbon KLL peak, and oxygen KLL peak to quantify the elemental composition and surface coverage with carbon 
and oxygen. From the Cu(MVV) and Al, Al$_2$O$_3$(LMM) peaks, we determine that the surface metal composition had close to ${50\,{\rm at.}\%}$ of copper and aluminum throughout our study. 
We show the results of the quantitative analysis for carbon and oxygen coverage in Fig.~\ref{data}b. The  first set of values, A$_{(1)}$-D$_{(1)}$,  corresponds to a generic energy-dependent 
sensitivity of our electron analyzer. The second set of values corresponds to an empirically determined sensitivity, which gives more self-consistent surface coverage information (see App.~
\ref{app:B}). 

In parallel with these measurements, we laser cooled single \Ca~ions and used the ion heating rate to extract the electric field noise spectral density at the ion frequency. We measured 
heating rates by laser cooling the ion, waiting for a variable delay time, and determining the change in the average population of one particular mode of the ion motion, denoted here by 
$\bar{n}$. This method determines the noise spectral density for one particular component of the electric field, in our case the electric field along the trap axis (direction 
$1//\sqrt{2}\,(-1,1,0)$ in Fig.~\ref{fig:sphere}). The electric field noise spectral density is given by $S_{E}= \frac{4\,m\,\hbar\,\omega}{e^2}\dot{\bar{n}}$, where $m$ is the ion mass, 
$\omega$ the frequency of the measured ion mode, and $e$ the elementary charge \cite{DesLauriers2006a}.

We used spectroscopy and resolved sideband cooling on the S$_{1/2}$-D$_{5/2}$ transition of the \Ca~ion to determine the ion temperature, and thus the heating rate. Before cleaning the 
surface of our trap, the high ion heating rates prevented us from sideband cooling to the motional ground state, thus the heating rates were measured on ions cooled to the Doppler limit 
of the S$_{1/2}$-P$_{1/2}$ transition. Specifically, we determined the mean value of the thermal state of the ion motion by measuring the collapse of Rabi oscillations on the carrier and 
sideband transitions \cite{Wineland1998}. After surface cleaning, the low noise of the trap allowed us to measure the heating rates on ions cooled to close to their ground state, by 
monitoring the relative strength of the red and blue sidebands of the S$_{1/2}$-D$_{5/2}$ transition \cite{Wineland1998}. We confirmed the consistency of these two methods by also measuring 
heating rates for Doppler-cooled ions for the trap after cleaning.

Fig.~\ref{data}c shows the heating rates we obtained for the trap. Before cleaning the trap surface, we measured heating rates between 5000(1000) quanta/s and 200(10) quanta/s at motional 
frequencies between 246~kHz and 852~kHz (red points). The frequency scaling of these heating rates is $\dot{\bar{n}} \sim f^{-2.27(0.23)}$, which implies a frequency scaling 
$S_{E} \sim f^{-1.27(0.23)}$ for the noise. After cleaning the trap, we find much reduced heating rates between 
30(1) quanta/s and 3.8(0.5) quanta/s. The ion heating rates measured 2 days after cleaning the trap (green curve) are consistent with the values measured 40 days after cleaning (black curve), 
despite the difference in carbon and oxygen content of the surface, seen in the Auger spectra. This indicates that copper/aluminum surfaces do not need to be atomically clean and oxide-free 
to achieve the low electric field noise levels in our trap.

For the cleaned trap, we also observe a change of the spectral characteristics of the noise in the frequency range between 200~kHz and 1~MHz. The heating rates decrease with frequency between 
200~kHz and 580~kHz, and at higher frequencies they show a broad maximum centered roughly around 800~kHz. The drop-off with frequency below 580~kHz has scaling $\dot{\bar{n}}\sim f^{-0.95(0.28)}$, 
implying $S_{E} \sim f^{-0.95(0.28)}$. This scaling is consistent with what has been measured elsewhere \cite{Labaziewicz2008a, Allcock2011}. However,  the measured noise levels are close to 
the values expected from Johnson noise in our system, and we suspect that at this point we are limited by technical noise in our system, rather than surface-related processes. The expected 
Johnson noise level is determined by the wirebonds and the capacitors in our setup, corresponding to 6-8 $\Omega$ for the real part of the impedance (blue line in Fig.~\ref{data}c). The peak 
at higher frequencies is reminiscent of resonant behavior and is consistent with a resonance in our filter board for certain parameter values.

\begin{figure}
\begin{center}
\includegraphics[width = 0.45\textwidth]{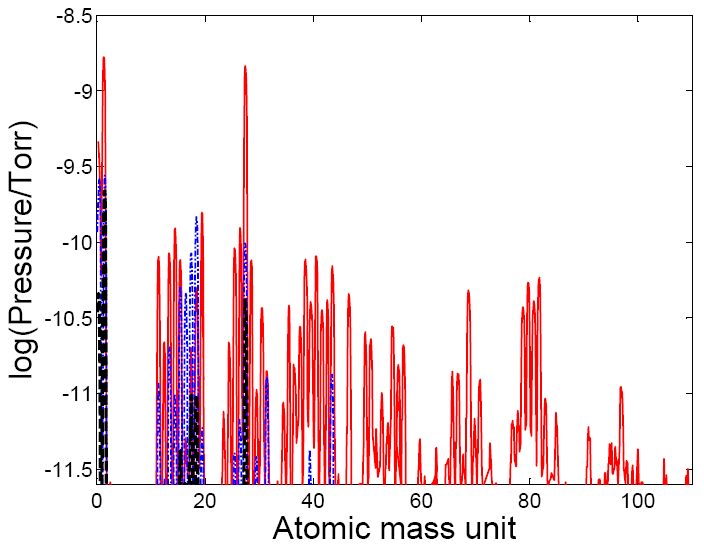}
\caption{Mass spectra taken with the residual gas analyzer (RGA). The spectrum taken immediately upon activating the RGA after baking (red) reveals that heavy hydrocarbon chains were 
deposited on the RGA surfaces during the bake. The black (dashed) spectrum, taken after 15 hours of operation of the RGA when its surfaces were clean, shows residual gases in the vacuum 
system after the bake. The  blue spectrum, taken immediately upon activating the RGA after it has been off for 60 days in ultra-high vacuum, shows that heavy hydrocarbons do not redeposit 
without baking (color online).} \label{rgadata}
\end{center}
\end{figure}

The mass spectra taken with the residual gas analyzer during the first few minutes of activating it, provide complementary information on the contaminants present in our vacuum system. 
Such spectra are a result of outgassing of the heated surfaces of the RGA (constructed from stainless steel, iridium, platinum, and ceramics), and they indicate what types of adsorbates 
reside on these surfaces. In Fig.~\ref{rgadata} we show in red a spectrum recorded immediately upon switching on the RGA  after the vacuum bake of our system (the RGA had been thoroughly 
outgassed prior to baking). It shows a wide range of heavy hydrocarbon chains, with weights up to 100~amu. After running the RGA for several hours, the spectra became cleaner (spectrum 
in black), showing that H$_2$, and hydrocarbons around CH$_4$ and CO were the predominant residual gases in our vacuum system. Subsequently, the RGA was switched off for 60 days in ultra-high 
vacuum. The  blue spectrum was taken immediately upon activating it after 60 days, and is free of heavy hydrocarbon chains. Thus, the vacuum bake resulted in deposition of heavy hydrocarbon 
chains on the surfaces of our RGA, while subsequent deposition is free of those. This suggests that the size of carbon compounds which contaminate the aluminum-copper surface of our trap 
could also be different in the different steps presented in Fig.~\ref{data}.

\begin{figure}
\begin{center}
\includegraphics[width = 0.48\textwidth]{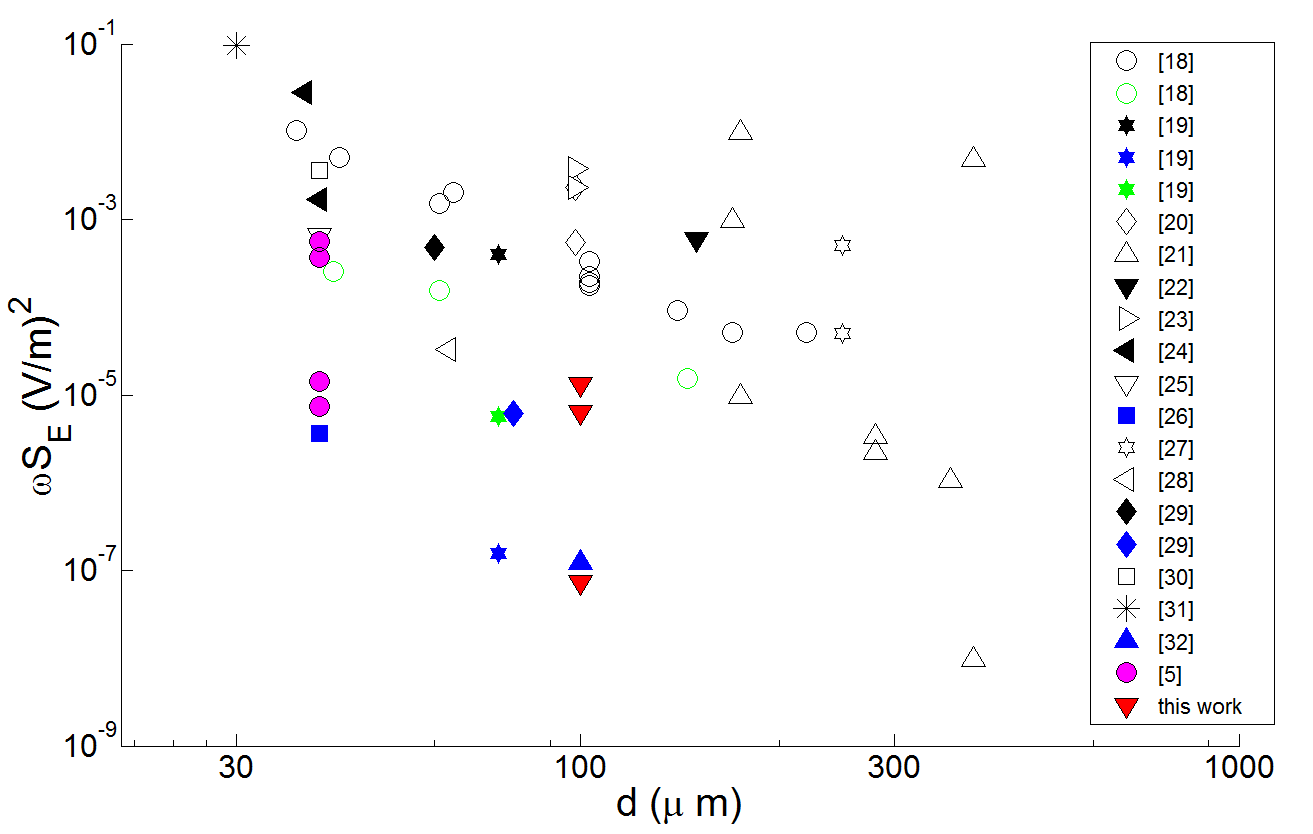}
\caption{Summary of representative electric field noise measurements for various ion traps. Measurements done at room temperature are shown in black, those done below room temperature in 
green (150~K$>$T$>$100~K), and blue (T$<$15~K). For the noise measurements reported here and in \cite{Hite2012} the non-$f^{-1}$ noise scaling results in a scatter of the measured noise 
spectra on an $\omega\,S_{\rm E}$ plot. The size of this range is indicated using multiple data points for the data showing $f^{-\alpha}$ scaling (color online).}\label{globalplot}
\end{center}
\end{figure}

It is instructive to compare the electric field noise measured in this work with noise measured on other traps. To partially take into account the $f^{-\alpha}$ noise spectrum, it is customary 
to compare $\omega\,S_{\rm E}$ for different traps. In Fig.~\ref{globalplot} we summarize $\omega\,S_{\rm E}$ versus the closest ion-electrode distance, for a number of trap noise measurements 
reported in the literature \cite{Hite2012,Turchette2000,Allcock2010,Allcock2011,Allcock2011a,Amini2010,Britton2009, Brown2011,Daniilidis2011,DesLauriers2006a, Doret2012,Labaziewicz2008a,Leibrandt2009,Seidelin2006,Stick2006,Wang2010}. 
To compare between ion traps with different dimensions, one scales the noise with ion distance from the surface, $d$, as $S_{\rm E}\sim d^{-4}$. This scaling is expected for noise arising from 
independently fluctuating, electrical-dipole like sources on the surface \cite{Turchette2000}. Taking this type of scaling into account, our noise measurements prior to cleaning fall roughly 
a factor of 10 below the general trend of room temperature noise measurements. After cleaning, the noise of our trap is comparable to the best cryogenic traps. These observations suggest that 
aluminum and copper can be good candidates as trap fabrication materials.

Our results give insight into the sources of noise near metal surfaces in vacuum. The RGA spectra measured after baking suggest that the vacuum bake resulted in coverage of 
some of the chamber surfaces with heavy hydrocarbon chains. The Auger spectra from the trap surface show that the initial coverage after baking contained higher fractions of carbon, whereas 
post-cleaning, the trap surface contained more oxygen. After cleaning, the noise levels on an aluminum-copper surface which still had traces of carbon and oxygen dropped significantly. We 
also found that small variations in carbon and oxygen did not affect the noise at our level of sensitivity. These observations suggest that the dominant contribution of the noise coming from 
metal surfaces in vacuum can critically depend on factors such as the size and chemical composition of the adsorbed hydrocarbon molecules, and the precise surface coverage. Such different 
mechanisms should be distinguishable experimentally by introducing controlled amounts of specific adsorbates to cleaned surfaces, and more work is required to fully understand these effects.

In summary, we have probed the electric field noise near an aluminum-copper surface at room temperature using a single trapped ion, and performed \emph{in-situ} elemental analysis and 
treatment of the surface. We find that cleaning of the surface using energetic argon ions reduces the noise to levels comparable to those of the best cryogenic traps and that the surface 
need not be atomically clean to show such low noise levels. Our findings further indicate that the type of contaminant species and the exact surface coverage fraction might play an important 
role in the electric field noise of surfaces.

We would like to acknowledge M. Crommie for useful feedback, and the loan of a LEED/AES spectrometer, and D. Hite, D. Leibfried, G. Somorjai, E. Yablanovitch, and J. Bokor for useful 
discussions. This research was funded by the Office of the Director of National Intelligence (ODNI), Intelligence Advanced Research Projects Activity (IARPA), through the Army Research 
Office grant W911NF-10-1-0284. All statements of fact, opinion or conclusions contained herein are those of the authors and should not be construed as representing the official views or 
policies of IARPA, the ODNI, or the U.S. Government.

\newpage
\appendix

\section{Suppression of electronic noise} \label{app:A}

The static voltages for trapping are generated by low-noise digital-to-analog converters (AD660). To reduce electronic noise injected to the trap electrodes, we
filter the static voltage signals using sixth-order low-pass filters with insertion loss higher than 90 dB in the frequency range between 200~kHz and 1~MHz. 
The last filtering stage resides inside vacuum, and consists of 47~nF (AVX, X7R dielectric) capacitors on the printed circuit board which supports the trap chip 
carrier, as well as 0.6~nF (AVX, X7S dielectric) capacitors, which are wire-bonded on the chip carrier. The entire vacuum system, including the electronics for 
the radio-frequency and static trapping potentials is housed inside a Faraday cage providing more than 40~dB of attenuation for electromagnetic fields in the range 
of frequencies between 200~kHz and 1~MHz, in which we have performed electric field noise measurements.

\section{Analysis of the Auger spectra} \label{app:B}

Our Auger electron analyzer (OCI microengineering, BLD-8000, with hemispherical retarding field analyzer) allowed us to reliably extract qualitative 
information about the surface composition, but only imprecise quantitative information. To quantitatively analyze the carbon and oxygen content, we 
have to correct for the difference in the energy-dependent sensitivity between our electron analyzer and cylindrical mirror analyzers for which Auger 
peak intensities are tabulated \cite{Davis1978}. In Fig.~2b we show a set of values (A$_{(1)}$-D$_{(1)}$) obtained by rescaling the peak 
intensities by a factor proportional to the peak energy, $E$, as expected for a generic retarding-field analyzer \cite{Davis1978}. However, the coverage 
values obtained with this rescaling are inconsistent with the observed changes of the Cu and Al, Al$_2$O$_3$ peak heights \cite{Davis1978}. As an 
alternative, we rescale the peak intensities by factors of the form $E+c\,E^n,\,n=2,3,...$. We find that rescaling by a factor 
$E/\rm{eV}$+1.3$\times10^{-6}(E/ \rm{eV} )^4$ yields more self-consistent results. We display these results in (A$_{(2)}$-D$_{(2)}$) of Fig.~2b.

\bibliographystyle{unsrt}

%\bibliography{library.bib}

\begin{thebibliography}{10}

\bibitem{Simoen2012}
E.~Simoen, Maria Gl\'{o}ria Ca\~{n}o de~Andrade, M.~Aoulaiche, N.~Collaert, and Cor Claeys.
\newblock {\em IEEE Transactions on Electron Devices}, 59(5):1272--1278, May 2012.

\bibitem{Ishigami2006}
Masa Ishigami, J.~H. Chen, E.~D. Williams, David Tobias, Y.~F. Chen, and M.~S. Fuhrer.
\newblock {\em Applied Physics Letters}, 88(20):203116, 2006.

\bibitem{Pashkin2009}
Yu.~A. Pashkin, O.~Astafiev, T.~Yamamoto, Y.~Nakamura, and J.~S. Tsai.
\newblock {\em Quantum Information Processing}, 8(2-3):55--80, 2009.

\bibitem{Stipe2001}
B~C Stipe, H~J Mamin, T~D Stowe, T~W Kenny, and D~Rugar.
\newblock {\em Physical Review Letters}, 87(9):096801, 2001.

\bibitem{Hite2012}
D~A Hite, Y~Colombe, A~C Wilson, K~R Brown, U~Warring, J~J\"{o}rdens, J~D Jost, K~S McKay, D~P Pappas, D~Leibfried, and D~J Wineland.
\newblock {\em Phys Rev Lett}, 109:103001, 2012.

\bibitem{Henkel1999}
C.~Henkel, S.~P\"{o}tting, and M.~Wilkens.
\newblock {\em Applied Physics B: Lasers and Optics}, 69(5-6):379--387, 1999.

\bibitem{Kim2010a}
W.~J. Kim, A.~O. Sushkov, D.~A.~R. Dalvit, and S.~K. Lamoreaux.
\newblock {\em Physical Review A}, 81(2):022505, 2010.

\bibitem{Pollack2008}
S.~Pollack, S.~Schlamminger, and J.~.H~Gundlach.
\newblock {\em Physical Review Letters}, 101(7):071101, 2008.

\bibitem{Everitt2011}
C.~W.~F. Everitt \emph{ et al.}
\newblock {\em Physical Review Letters}, 106:221101, 2011.

\bibitem{Wineland1998}
D~J Wineland, C~Monroe, W~M Itano, D~Leibfried, B~E King, and D~M Meekhof.
\newblock {\em Journal of Research of the National Institute for Standards and Technology}, 103:259--328, 1998.

\bibitem{Clerk2010}
A.~A. Clerk, M.~H. Devoret, S.~M. Girvin, Florian Marquardt, and R.~J.
  Schoelkopf.
\newblock {\em Reviews of Modern Physics}, 82(2):1155--1208, 2010.

\bibitem{Teufel2009}
J~D Teufel, T~Donner, M~A Castellanos-Beltran, J~W Harlow, and K~W Lehnert.
\newblock {\em Nature nanotechnology}, 4(12):820--3, 2009.

\bibitem{Maiwald2009a}
R~Maiwald, D~Leibfried, J~Britton, J.~C. Bergquist, G~Leuchs, and D.~J.
  Wineland.
\newblock {\em Nature Physics}, 5(8):551--554, 2009.

\bibitem{Budker2007}
D~Budker and M~Romalis.
\newblock {\em Nature Physics}, 3:227--234, 2007.

\bibitem{Devoret2000}
M~H~Devoret and R~J~Schoelkopf.
\newblock {\em Nature}, 406(6799):1039--46, 2000.

\bibitem{Haeffner2008}
H.~H\"{a}ffner, C.~F. Roos, and R.~Blatt.
\newblock {\em Physics Reports}, 469:155, 2008.

\bibitem{Davis1978}
L.~E. Davis, N.~C. MacDonald, P.~W. Palmberg, G.~E. Riach, and R.~E. Weber.
\newblock {\em {Handbook of Auger Electron Spectroscopy}}.
\newblock {Perkin-Elmer, second edition, 1978}.
\newblock {\em Practical Surface Analysis}.
\newblock {John Wiley \& Sons, 1983}.

\bibitem{DesLauriers2006a}
L~Deslauriers, S~Olmschenk, D~Stick, W~K Hensinger, J~Sterk, and C~Monroe.
\newblock {\em Physical Review Letters}, 97(10):103007, 2006.

\bibitem{Labaziewicz2008a}
Jaroslaw Labaziewicz, Yufei Ge, David~R. Leibrandt, Shannon~X Wang, Ruth Shewmon,
  and Isaac~L Chuang.
\newblock {\em Physical Review Letters}, 101:180602, 2008.

\bibitem{Allcock2011}
D~T~C Allcock, L~Guidoni, T~P Harty, C~J Ballance, M~G Blain, a~M Steane, and
  D~M Lucas.
\newblock {\em New Journal of Physics}, 13(12):123023, 2011.

\bibitem{Turchette2000}
Q~A Turchette,D~Kielpinski, B~E King, D~Leibfried, D~M Meekhof, C~J Myatt, M~A
  Rowe, C~A Sackett, C~S Wood, W~M Itano, C~Monroe, and D~J Wineland.
\newblock {\em Physical Review A}, 61:063418, 2000.

\bibitem{Allcock2010}
D~T~C Allcock, J~A Sherman, D~N Stacey, A~H Burrell, M~J Curtis, G~Imreh, N~M
  Linke, D~J Szwer, S~C Webster, A~M Steane, and D~M Lucas.
\newblock {\em New Journal of Physics}, 12(5):053026, May 2010.

\bibitem{Allcock2011a}
D.~T.~C. Allcock, T.~P. Harty, H.~a. Janacek, N.~M. Linke, C.~J. Ballance,
  A.~M. Steane, D.~M. Lucas, R.~L. Jarecki, S.~D. Habermehl, M.~G. Blain,
  D.~Stick, and D.~L. Moehring.
\newblock {\em Applied Physics B}, 107(4):913--919, November 2011.

\bibitem{Amini2010}
J~M Amini, H~Uys, J~H Wesenberg, S~Seidelin, J~Britton, J~J Bollinger,
  D~Leibfried, C~Ospelkaus, a~P VanDevender, and D~J Wineland.
\newblock {\em New Journal of Physics}, 12(3):033031, March 2010.

\bibitem{Britton2009}
J.~Britton, D.~Leibfried, J.~a. Beall, R.~B. Blakestad, J.~H. Wesenberg, and
  D.~J. Wineland.
\newblock {\em Applied Physics Letters}, 95(17):173102, 2009.

\bibitem{Brown2011}
K~R Brown, C~Ospelkaus, Y~Colombe, a~C Wilson, D~Leibfried, and D~J Wineland.
\newblock {\em Nature}, 471(7337):196--9, March 2011.

\bibitem{Daniilidis2011}
N~Daniilidis, S~Narayanan, S~M\"{o}ller, R~Clark, T~Lee, P~Leek, A~Wallraff,
  St~Schulz, F~Schmidt-Kaler, and H~H\"{a}ffner.
\newblock {\em New J. Phys.}, 13:013032, 2011.

\bibitem{Doret2012}
S~{Charles Doret}, Jason~M Amini, Kenneth Wright, Curtis Volin, Tyler Killian,
  Arkadas Ozakin, Douglas Denison, Harley Hayden, C-S Pai, Richart~E Slusher,
  and Alexa~W Harter.
\newblock {\em New Journal of Physics}, 14(7):073012, July 2012.

\bibitem{Leibrandt2009}
D~R Leibrandt, J~Labaziewicz, R~J Clark, I~L Chuang, R~Epstein, C~Ospelkaus,
  J~Wesenberg, J~Bollinger, D~Leibfried, D~Wineland, D~Stick, J~Sterk,
  C~Monroe, C.-S. Pai, Y~Low, R~Frahm, and R~E Slusher.
\newblock {\em Quantum Information and Computation}, 9(11):0901, 2009.

\bibitem{Seidelin2006}
S~Seidelin, J~Chiaverini, R~Reichle, J~J Bollinger, D~Leibfried, J~Britton, J~H
  Wesenberg, R~B Blakestad, R~J Epstein, D~B Hume, W~M Itano, J~D Jost,
  C~Langer, R~Ozeri, N~Shiga, and D~J Wineland.
\newblock {\em Physical Review Letters}, 96:253003, 2006.

\bibitem{Stick2006}
D~Stick, W~K Hensinger, S~Olmschenk, M~J Madsen, K~Schwab, and C~Monroe.
\newblock {\em Nature Physics}, 2:36, 2006.

\bibitem{Wang2010}
Shannon~X. Wang, Yufei Ge, Jaroslaw Labaziewicz, Eric Dauler, Karl Berggren,
  and Isaac~L. Chuang.
\newblock {\em Applied Physics Letters}, 97(24):244102, 2010.

\end{thebibliography}

\end{document}